\documentclass[10pt,a4paper]{article}

\usepackage{amscd} \usepackage{amsmath} \usepackage{amsfonts}
\usepackage{graphics}\usepackage{epic}
\usepackage{pstricks} \usepackage{latexsym}

\def\be{\begin{equation}}
\def\ee{\end{equation}}
\def\bea{\begin{eqnarray}}
\def\eea{\end{eqnarray}}
\def\d{\partial}

\def\A {$AdS_3$}
\def\AA {$AdS_2$}
\def \SL {$SL_2(\mathbb{R})$}

\def \H {$H_3^+$}

\def \Z {\mathbb{Z}}

\def \R {\mathbb{R}}
\def \N {\mathbb{N}}

\def \a {\alpha'}
\def \sl {$\widehat{sl(2)}$}

\title{{\small \begin{flushright}
CPHT-062.0702 \\ \vskip .3cm
\end{flushright}} \AA\ D-branes in \A\ spacetime}
\author{Sylvain Ribault\footnote{ribault@cpht.polytechnique.fr }\\
{\it\small  Centre de Physique Th{\'e}orique, Ecole Polytechnique,
91128 Palaiseau, FRANCE}\\
{\small Based on a talk at the 2001 Corfu Summer Institute on Elementary
Particle Physics}}
\date{}

\begin{document}





\maketitle

\centerline{ \bf Abstract}

\vskip .2cm

I review some recent progress in understanding the properties
of \AA\ branes in \A. Different methods -- classical string
motion, Born-Infeld dynamics, boundary states -- are evocated and
compared.

\section{Introduction}

A prominent motivation for the study of strings in \A\ is
the AdS/CFT correspondence \cite{ori,rev}. This is a holographic
relationship between 
string theory in an Anti-de Sitter
 spacetime and field theory on its conformal
boundary. The relationship also extends to D-branes in \A. This has
been made recently quite precise in the case of the \AA\ D-brane
\cite{bbdo,asy}. The \AA\ brane extends to the boundary of \A\ where
it may be interpreted as a one-dimensional conformal wall. 
 
I will not focus on this holographic interpretation here, but rather on
the properties of the \AA\ branes themselves~: their geometry, 
open-string spectrum, and interactions with closed strings. I do not
present any new material, but summarize and relate recent results. 

Let me briefly describe the geometry of \A. Define it as the following
hypersurface of $\R^{2,2}$~:
$-X_0^2-X_1^2+X^2_2+X^2_3=-L^2$. This is a solution to the Einstein
equations with negative cosmological constant. This solution has
constant negative curvature. The link with the group \SL\ is made through 
\bea
g = \frac{1}{L}\
\left(
\begin{array}{lll}
X^0+X^1 &\quad & X^2+X^3\cr
X^2-X^3 &\quad & X^0- X^1\cr
\end{array}
\right)  
\eea
There is another useful parametrization, the global coordinates~:
\bea
\left\{
\begin{array}{lll}
X_2+iX_3 &=& L\sinh\rho \ e^{i\theta} \cr
X_0+iX_1 &=& L\cosh\rho \ e^{i\tau} \cr
\end{array}
\right.
\eea
In these coordinates \A\ is a solid cylinder ($\tau$ takes all real
values) and \SL\ is a solid torus ($\tau$ is compactified). The metric
is 
\bea ds^2=L^2(-\cosh^2\rho\
d\tau^2+d\rho^2+\sinh^2\rho\ d\theta^2)=-dX_0^2-dX_1^2+dX^2_2+dX^2_3. \eea

\A\ is also part of a string background. This can be seen at the level
of supergravity~: a solution can be written using the metric of \A\ and
an NS-NS three-form $H$
(the volume form of \A). In the equation of motion for the metric,
there is no more cosmological constant, but instead the square of $H$.

There are branes with induced \AA\ geometry
in the \A\ background \cite{bp,stan}. They
 have the following equations in global
coordinates~:
\bea \sinh\rho\ \cos\theta=\sinh \psi \label{geom}  \eea
The constant $\psi$ is in fact quantized, as we will see later. Now a
D-brane is really defined by the geometry, plus the value of the
worldvolume two-form $F$. This depends on a choice of gauge for the
$B$-field such that $H=dB$. There is a gauge in which we have on each
brane~:
\bea B+2\pi\a F=\frac{\tanh\psi}{L^2}\ vol\ ,\ \ 2\pi\a
F=-\frac{\psi}{L^2\cosh^2 \psi}\ vol\ , \label{ffield}  \eea
where $vol$ is the volume form induced on the brane by the bulk
metric.    

We have precisely defined the branes we are interested in. The main tool
for studying these branes will be the theory of strings and branes in group
manifolds. The main difficulties in this case come from the
non-compactness of \A\ (this results in volume divergences in some physical
quantities) and its Minkowskian signature.

\section{D-branes in the \SL\ WZW model}

The \AA\ branes have the general properties of twined conjugacy classes
in WZW
models. These branes have been known for some time \cite{fffs}. 
Let me call
 $g(z,\bar{z})$ the embedding of the string worldsheet into a group manifold.
Taking advantage of the group structure, 
we construct right- and
left-moving currents
\[ g(z,\bar{z}) \rightarrow J_L=-g^{-1}\partial g\ ;\
J_R=\bar{\partial}gg^{-1} \]
From the modes of these currents (i.e. the coefficients of their
decomposition in powers of $z$ and $\bar{z}$) we construct
two copies of the affine Lie algebra of the group, which
are symmetries of the
closed-string spectrum.

Let $\omega$ be an automorphism of the Lie algebra of our group. We
use the same notation for its natural extension to the affine Lie algebra.
The boundary conditions: $J_L=\omega(J_R)$, preserve the conformal
symmetry. If $\omega$ is an outer automorphism they break the Lie
algebra symmetry and the corresponding brane has the geometry of a twined
conjugacy class~: $\{hg\omega(h)^{-1}\}$, for $g$ fixed and $h$
varying on the group. 

The group \SL\ has an outer automorphism~: $g\rightarrow \Omega g
\Omega$, where $\Omega =  
{\tiny \left(
\begin{array}{ll}
 0 & 1 \cr
 1 & 0 \cr
\end{array}
\right)} $. The corresponding twined conjugacy classes have equations
 of the following form~:
 ${\rm Trace\ } \Omega g =\ {\rm cst}$.
This amounts to eq. (\ref{geom}) if we write ${\rm cst}\ =2\sinh \psi$. General
expressions are also known for the $B$ and $F$ fields \cite{brs},
which in our case amount to eq. (\ref{ffield}).

In compact groups this comes with a quantization of the allowed
positions of the D-branes due to the mechanism of flux stabilization
\cite{flux,brs}. This is due to the nontrivial topology of the
D-brane, more precisely to its non-trivial two-cycles. Our \AA\ branes
have a trivial topology but their position is still quantized
\cite{bp}~: $\sinh\psi =qT_F/T_D$ where $q\in\Z$ and $T_D,T_F$ are the
D-string and fundamental string tensions.

Now let us list some interesting physical quantities characterizing
D-branes. The exact quantities are defined in the worldsheet approach,
i.e. 
from boundary conformal field theory. An approximate space-time
description is given by the Born-Infeld action.
\[
\begin{array}{r|l}
\rm{SPACE-TIME} & \rm{WORLD-SHEET} \\ \hline \\
S^{BI}(p,F)+\rm{corrections}
&  S^{\sigma}=\frac{1}{4\pi\a}\int _{\Sigma} g_{\mu \nu}+B_{\mu
  \nu}+\int _{\partial \Sigma} A \\
\rm {Born-Infeld\ solution} & \rm{boundary\ state} \\
\rm {position} & \rm{v.e.v.\ of\ graviton\ or\ dilaton} \\
\rm {energy} & \rm {quantum\ dimension\ (mass)} \\ 
\rm {quadratic\ fluctuations} & \rm {open-string\ spectrum} \\

\end{array}
\]
\begin{itemize}
\item $\sigma$ denotes the sigma-model, $S^\sigma$ is the
  Polyakov action.
$\Sigma $ is the worldsheet and its boundary $\partial\Sigma$ lies
  in the D-brane. 
\item $S^{BI}=\int \sqrt{\det(p^*g_{\mu\nu}+p^*B_{\mu\nu}+2\pi
    \alpha'F)}$ 
is the Born-Infeld action, depending on the embedding
  $p$ of the D-brane in space-time and on the two-form field $F=dA$ on the
  D-brane. It is the first term in an expansion in powers of $\a$ (not
  counting the 
  factor $\a$ of $F$). In the case of group manifolds, the metric and
  $B$ field are
  proportional to $L^2=k\a$ where $k$ is called the level. So the
  expansion may be written in powers of $1/k$.
\item A D-brane, solution of the Born-Infeld equations of motion, is
  described in CFT by its interactions with closed strings, i.e. by a
  linear form on closed strings states, i.e. by a closed string state
  called the boundary state. In particular, the interactions with
  gravitons and dilatons
betray the position of the brane. This is explicitly seen
  in \cite{fffs}. 
\item The correspondence between the space-time energy and the quantum
  dimension is explained in \cite{tid}.
\item The spectrum of Born-Infeld fluctuations around a D-brane is the
  point-particle limit ($k\rightarrow \infty$) of the open-string
  spectrum. This can be seen explicitly in the case of compact groups
  \cite{flux, brs}.
\end{itemize}
In our case of the \AA\ brane, the space-time quantities are easily
computed \cite{bp,pr}. We already described the position and F-field
of the brane. Its energy is proportional to $\cosh \psi$ (for a
$\psi$-independent regularization of the volume divergence). The
spectrum of fluctuations is basically the space of functions on the
\AA; the Born-Infeld equations of motion should really be solved only
in a complete (ten-dimensional) string background, and they roughly 
correspond
to the Virasoro conditions.

To describe this spectrum we organize it in representations of \SL\
(which acts on the brane by conjugation). It is time to describe those
representations as well as the unitary representations of the
corresponding affine Lie algebra \sl.

\section{ Closed and open strings in \A}

Let me first review some material from \cite{moi}.
We describe states in representations of \SL\ by the spin $j$
(corresponding to a Casimir eigenvalue $-j(j-1)$) and the eigenvalue
$m$ of some generator $J^3$ of the Lie algebra. We will use discrete
representations $D_j^\pm$ 
with $j>0\in \R$ and $m\in \pm j\pm \N$~; and continuous representations
$C_j^\alpha$ with $j\in \frac{1}{2}+i\R$ and $0 \leq\alpha<1$ and
$m\in \alpha+\Z$. Then one can build 
representations of \sl\ as follows~: \sl\ can be defined by
its generators $J^a_n$ (where $n\in\Z$ and $a$ is an $sl(2,\R)$ label),
and the relations
\bea
[J^a_n,J^b_m]=f^{ab}_cJ^c_{m+n}+k\frac{n}{2}\kappa^{ab}\delta{m,-n},\eea
where $f^{ab}_c$ and $\kappa ^{ab}$ are the $sl(2,\R)$ structure
constants and Killing form. Unflowed representations of \sl\ are
defined by starting with an $sl(2,\R)$ representation $C_j^\alpha$ (or
$D_j^\pm$)
and acting on it
with the operators $J^a_{-n}$ (with $n>0$~; the action of operators
$J^a_n$ is assumed to give zero) to get the affine 
$\hat{C}_j^\alpha$ (or $\hat{D}_j^\pm $) representation. 
Flowed representation can
then be built using the spectral flow automorphisms of \sl, defined
for any $w\in \Z$~:
\bea J^3_n,\ J^\pm_n\ \longrightarrow\ J^3_n-k\frac{w}{2}\delta_{n,0},\
J^\pm_{n\pm w}. \eea
(we use labels $a=3,+,-$ corresponding to the $sl(2,\R)$ relations
$[J^3,J^\pm]=\pm J^\pm,\ [J^+,J^-]=2J^3$). They are called
$\hat{C}_j^{\alpha,w},\hat{D}_j^{\pm,w} $ .

This is enough to build the spectrum of closed strings in \A~: it is
made of the above-mentioned representations with the constraint
$\frac{1}{2}<j<\frac{k-1}{2}$ on the discrete representations, 
each tensored with a
right-moving copy of itself. This spectrum can be checked by exact
partition function computations in \H \cite{gaw,moii} (the Euclidean
version of \A, 
obtained by $\tau\rightarrow i\tau$). One can also build consistent
three- and four-point correlators in \H\cite{jt,moiii}.

One way to have an intuition of these different kinds of
representations is to study classical closed string motion in \SL. The
continuous representations correspond to strings reaching the spatial
infinity (long strings), whereas discrete representations correspond to strings
staying at finite distance (short strings) \cite{moi}. We can also do
such an analysis 
for open strings ending on an \AA\ brane \cite{pr,plot}.
Let me call  $x^\pm=\sigma^0\pm\sigma^1$ the world-sheet light-cone 
coordinates
($\sigma^1 \in [0,\pi]$) and
$J_+=-\d_+gg^{-1},\ J_-= g^{-1}\d_-g$ the world-sheet currents.
The general solution to the bulk equations of motion, plus boundary
condition $J_-=\omega(J_+)$ at $\sigma ^1=0$ is \cite{cla,pr}
\bea g=a(x^+)\, m\, \omega( a(x^-))^{-1}, \eea
where $a$ is a group-valued function which should be $2\pi$-periodic
up to left-translation by a constant, and $m$ is a group element (defining
the twined conjugacy class to which $\sigma ^1=0$ is attached). 

To have an idea of the different sectors appearing in the open-string
spectrum it is enough to first restrict ourselves to functions
$a(x)=e ^{xC}$, where $C$ is a constant matrix, then apply
spectral flow $a\rightarrow \exp xw\left(
\begin{array}{ll}
 0 & 1/2 \cr
 -1/2 & 0 \cr
\end{array}
\right)\ \times a$. However, if $w$ is odd, then the image of our open
string has no more its two ends on the same D-brane, but stretches
between two opposite branes of parameters $\psi,-\psi$. If we
restrict to the study of a single brane, it is still possible to
construct classical solutions which will be interpreted as strings in
$\hat{D}_j^{\pm,w}$, but this no longer works for long strings
\cite{plot}. For those, only half of the spectral flow symmetry is preserved.

In the case of the "straight brane" $\psi=0$, such problems do not
arise. Moreover, all classical open string solutions give rise to
closed string solutions by reflection with respect to the brane\cite{pr}. It is
then natural to expect that the open-string spectrum is the
holomorphic square-root of the closed-string spectrum (i.e. it
consists of the same representations, but without the right-movers), 
as can indeed be proved \cite{plot}.

\section{Some speculations}

I finally comment on the significance of these results for the exact
spectrum of open strings living on the \AA\ brane in \A. 

Since the r{\^o}le of spectral flow seems quite clear, what remains to
be determined is the density of unflowed representations in the
spectrum. Let me start by coming back to the Born-Infeld analysis. It
predicts a constant density of discrete representations, independent
of the position $\psi$ of the brane. The continuous representations
are not really seen here because they have positive Casimir, thus they
are not able to fulfill the Virasoro conditions in a background whose
only time direction is included in \A. They nevertheless appear in the
physical spectrum because of spectral flow (which changes the Virasoro
generators), however the Born-Infeld analysis is valid in the
point-particle limit in which there is no spectral flow (because we are
really dealing with representations of the mere Lie algebra, not \sl).

In the case of the $S^2$ D-brane in $SU(2)$, the Born-Infeld analysis
is known to give exact results for the open-string density of states,
up to a truncation on the allowed spins \cite{flux}. This could be
due to the vanishing of contributions from corrections to the
Born-Infeld action, which would not be surprising given the
supersymmetry of the $AdS_2\times S^2$ D-brane in the 
background $AdS_3\times SU(2)\times U(1)^4$ \cite{bp,bac}. If these
corrections vanish for the $S^2$ brane in $SU(2)$ they may as well
vanish for the \AA\ brane in \A, given the formal similarity between
the two cases. This would indicate that the spectrum of discrete
representations 
on \AA\ branes is brane-independent.

We can also compare the \AA\ brane to its counterpart in \H, obtained
by the analytic continuation $\tau\rightarrow i\tau$. For this
Euclidean \AA\ brane in \H, exact expressions for the boundary state
and the open-string spectrum are known \cite{pst,plo} (however, there
are still problems to be solved 
in the construction of the corresponding boundary
CFT, see \cite{pon}). Because the \AA\ brane in \A\ is $\tau$-independent, it
is still possible to interpret these expressions for this brane. Only
continuous representations appear in the spectrum of \H, so the
boundary state for \AA\ in \H\ does not give a complete boundary state
in \A. This is not really a problem since the discrete closed string
states are not expected to couple to the \AA\ brane because it is
static (this can be seen in harmonic analysis~: the integral over \AA\
of a
point-particle wavefunction in a discrete representation
vanishes). However, discrete representations do appear in the
open-string spectrum of the \AA\ brane in \A, whereas they are absent
from the \H\ results. Here we should notice
that the Cardy computation in \cite{pst} takes
into account only the difference of open-string densities of states
between two D-branes of different positions. This means that this
computation is insensitive 
to $\psi$-independent terms in the density of states. To interpret
this in \A\ is thus possible, if the density of discrete
representations is brane-independent. This conclusion agrees with the
previous reasoning about effective actions.

We have argued 
that the density of states for open string living on a
single \AA\ brane in \A\ is brane-independent for discrete
representations, and given by \cite{pst} for continuous representations.

\vskip 0.2cm
\centerline{\bf Acknowledgements}
\vskip 0.1cm
\noindent
I thank the organizers of the Corfu meeting for the opportunity to
talk, Angelos Fotopoulos for carefully reading this manuscript, 
and Marios Petropoulos for discussions and encouragements.



\end{document}